\documentclass[twocolumn,preprintnumbers,amsmath,amssymb,superscriptaddress]{revtex4}

\usepackage{epsfig}

\newcommand{\ks}{K^{0}_{S}}
\newcommand{\ksp}{K^{0}_{S}\> p}
\newcommand{\kspb}{K^{0}_{S}\> \bar{p}}
\newcommand{\ksppb}{K^{0}_{S}\> p\>(\bar{p})}

\newcommand{\gev}{\; \mathrm{GeV}}
\newcommand{\mev}{\; \mathrm{MeV}}

\begin{document}

% Use the \preprint command to place your local institutional report
% number in the upper righthand corner of the title page in preprint mode.
% Multiple \preprint commands are allowed.
% Use the 'preprintnumbers' class option to override journal defaults
% to display numbers if necessary
\preprint{ANL-HEP-CP-04-34}

%Title of paper
\title{Observation of a narrow baryonic state in DIS at HERA}

% repeat the \author .. \affiliation  etc. as needed
% \email, \thanks, \homepage, \altaffiliation all apply to the current
% author. Explanatory text should go in the []'s, actual e-mail
% address or url should go in the {}'s for \email and \homepage.
% Please use the appropriate macro foreach each type of information

% \affiliation command applies to all authors since the last
% \affiliation command. The \affiliation command should follow the
% other information
% \affiliation can be followed by \email, \homepage, \thanks as well.
\author{S.V.Chekanov}
\email[]{chekanov@mail.desy.de}
%\homepage[]{Your web page}
\thanks{
Presented at the YITP workshop (YITP-W-03-21) "Multi-quark hadrons; four, five and more?", (February 17-19, 2004),  Kyoto, Japan. 
I would like to thank Prof. K.~Tokushuku for the financial support to come to this workshop.}
%\altaffiliation{}
\affiliation{HEP division, Argonne National Laboratory,
9700 S.Cass Avenue,
Argonne, IL 60439
USA}

%Collaboration name if desired (requires use of superscriptaddress
%option in \documentclass). \noaffiliation is required (may also be
%used with the \author command).
%\collaboration can be followed by \email, \homepage, \thanks as well.
\collaboration{for the ZEUS Collaboration}
%\noaffiliation

\date{\today}

\begin{abstract}
A resonance search has been made in the $\ksp$, $\kspb$ and $K^+
p$ invariant-mass spectra measured with the ZEUS detector at HERA
using an integrated luminosity of 121 pb$^{-1}$. The search was
performed in the central rapidity region of inclusive deep
inelastic scattering at an $ep$ centre-of-mass energy of 300--318
GeV for exchanged photon virtuality, $Q^2$, above 1 $\gev^2$. The
results support the existence of a narrow state in $\ksp$ and
$\kspb$  decay channels, consistent with the pentaquark prediction.
No signal was found in the $K^+ p$ decay
channel.
\end{abstract}

% insert suggested PACS numbers in braces on next line
% insert suggested keywords - APS authors don't need to do this
%\keywords{}

%\maketitle must follow title, authors, abstract, \pacs, and \keywords
\maketitle

% body of paper here - Use proper section commands
% References should be done using the \cite, \ref, and \label commands
\section{Introduction}

Recent results from fixed-target experiments triggered new
interest in baryon spectroscopy after observing a narrow baryonic 
resonance in the $K^+ n$ decay channel with a mass of
approximately $1530\mev$ and positive strangeness
\cite{prl91012003,prl91kub,prl92:032001,plb572:127}. Such a state can be
interpreted as a bound state of five quarks, i.e. as a pentaquark,
$\Theta^+ = uudd\bar{s}$. According to the predictions of the
chiral soliton model \cite{zp:a359:305}, 
$\ksp$ and $\kspb$ (denoted as
$\ksppb$) decays are also possible. 
For the $\ksp$ channel,  evidence for a corresponding
signal has been seen by other low-energy fixed-target experiments
\cite{pan66:1715,plb585:213,hep-ex-0309042,hep-ex-0401024,hep-ex-0403011}.
Recently, two other pentaquark candidates have  also been
reported \cite{prl92:042003,hep-ex-0403017}.

The $\ksppb$  decay  channel is
expected for the PDG  $\Sigma$ bumps \cite{pr:d66:010001},
which have been  observed by early fixed-target experiments.
Such unestablished
resonances are too low in mass to be accommodated in most quark models.
The presence of such states  in the mass regions close to the production
threshold of  $\ksppb$ final state
complicates the search for possible pentaquark.

The $\Theta^+$ state and the $\Sigma$ bumps discussed above have never
been observed in high-energy experiments, where hadron production is
dominated by fragmentation.

This paper discusses  
recent ZEUS results \cite{hep-ex-0403051} on the search for exotic
baryons in the $\ksppb$, $K^+p$, $K^-\bar{p}$ invariant-mass spectra
measured with the ZEUS detector at HERA. The measurements were  
based on the central pseudorapidity region ($|\eta|\leq 1.5$) where the contribution
from the fragmentation of the proton remnant is negligible. If the
pentaquark state is produced  without the net baryon number carried by
the proton remnant, emerging from the emissions of gluons and
quarks in the hadronisation process, this would open a new
chapter in our understanding of non-perturbative QCD.

%%%%%%%%%%%%%%%%%%%%%%%%%%%%%%%%%%%%%%%%%%%%%%%%%%%%%%%%%%%%%
\section{Evidence for  a baryonic state decaying to $\ksppb$}
%%%%%%%%%%%%%%%%%%%%%%%%%%%%%%%%%%%%%%%%%%%%%%%%%%%%%%%%%%%%%

The present analysis was performed using deep inelastic scattering
events measured with exchanged-photon virtuality $Q^2\ge 1\gev^2$.
The data sample corresponds to an integrated luminosity of 121
pb$^{-1}$, taken between 1996 and 2000. The analysis is based on
charged tracks measured in the central tracking (CTD). The tracks
were selected with $p_T\ge 0.15\gev$ and
$|\eta|\le 1.75$,
restricting this  study to a region where the CTD track acceptance
and resolution are high.

\begin{figure}
\begin{center}
  \includegraphics[height=6.0cm]{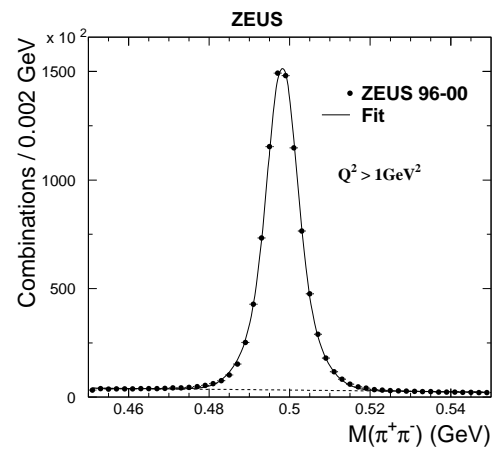}%
\caption{
The $\pi^+\pi^-$ invariant-mass distribution for $Q^2>1\gev^2$.
The solid line shows the fit result using two Gaussians  plus a
linear background, while the dashed line shows the linear background.
}
\label{pap0}
\end{center}
\end{figure}

\begin{figure}
\begin{center}
\includegraphics[height=6.0cm]{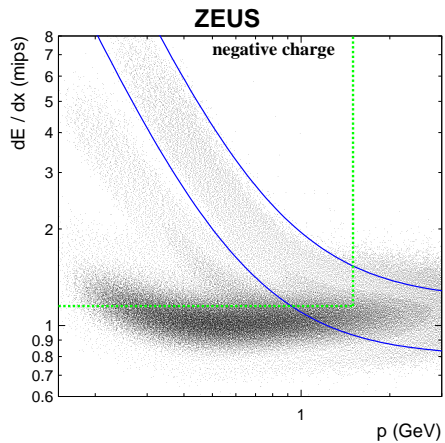}
\caption{ The $dE/dx$ distribution as a function of the track
momentum, $p$,  for negative tracks. The solid lines indicate the
antiproton band used in this analysis, while the dotted lines 
indicate the selection cuts $dE/dx > 1.15$ mips and $p < 1.5\gev$ used in  
the $\ksppb$ reconstruction.} \label{pap0dedx}
\end{center}
\end{figure}

$\ks$-mesons were identified by their charged-decay mode,
$\ks\to\pi^{+}\pi^{-}$ (see Fig.~\ref{pap0}). The candidates with
$p_{T}(\ks)>0.3\gev$ and $|\eta(\ks)|<1.5$ were retained. To
eliminate contamination from $\Lambda (\bar{\Lambda})$ 
decays, candidates with a
proton mass hypothesis $M(p\> \pi)<1121\mev$ were rejected.

The (anti)proton-candidate selection used the energy-loss
measurement in the CTD, $dE/dx$. As example, Fig.~\ref{pap0dedx}
shows the $dE/dx$ distribution as a function of the track momentum
for negative tracks. Clear antiproton and $K^-$-meson  bands were observed
($dE/dx$ for Monte Carlo simulations and for 
positive tracks have similar qualities, not shown).

$\ksppb$ invariant masses were obtained by combining $\ks$
candidates in the mass region $480-510\mev$ with (anti)proton
candidates in the (anti)proton $dE/dx$ band with the additional
requirements $p<1.5\gev$ and $dE/dx>1.15$ mips in order to reduce the
pion background.  The CTD resolution for the $\ksppb$ invariant-mass
near $1530\mev$,  estimated using Monte Carlo simulations, was  $2.0\pm
0.5\mev$ for both the $\ksp$ and
the $\kspb$ channels.

\begin{figure}
\includegraphics[height=8.0cm]{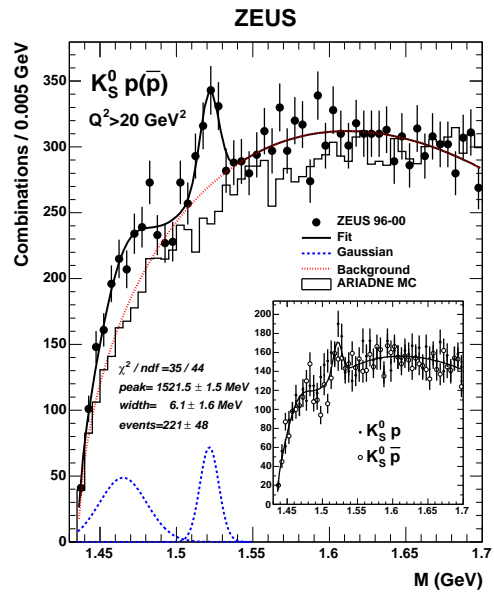}%
\caption{\label{theta}
Invariant-mass spectrum for the $\ksppb$ channel for $Q^2 >
20 \gev^2$. The solid line is the
result of a fit to the data using the threshold background
plus two Gaussians.  The dashed lines show the
Gaussian components, while the dotted line indicates background.
The prediction of the {\sc Ariadne} 
simulation is normalised to the data in the mass region above
$1650\mev$. The inset shows the $\kspb$ (open circles) and the $\ksp$
(black dots) candidates separately, compared to the result of the fit
to the combined sample scaled by a factor of 0.5.  }
\end{figure}

Figure~\ref{theta} shows the $\ksppb$ invariant mass  for $Q^2>
20\gev^2$, as well as for the $\ksp$ and $\kspb$ samples
separately (shown as inset). The distribution of the 
{\sc Ariadne} Monte Carlo model
was normalised to the data in the mass region above $1650\mev$.
The data are above the Monte Carlo prediction near  $1470\mev$ and
$1522\mev$, with a clear peak at
$1522\mev$.

To extract the signal seen at  $1522\mev$, the fit was performed
using a background function plus two Gaussians. The background has
the threshold form $F(M)=P_1(M-m_{K}-m_{p})^{P_2}(1+P_3(M-m_{K}-m_{p}))$,
where $m_K$ and $m_p$ are the masses of the kaon and the proton,
respectively, and $P_{i=1,2,3}$ are free parameters. 
The first Gaussian, which
significantly improves the fit at low masses, may  correspond to
the unestablished PDG $\Sigma (1480)$. The second Gaussian
describes a pronounced peak at $1522\mev$. The peak
position determined from the fit was $1521.5\pm 1.5({\rm
stat.})^{+2.8}_{-1.7} ({\rm syst.}) \mev$. 
It agrees well with the measurements by HERMES, SVD and COSY-TOF
for the same decay channel \cite{plb585:213,hep-ex-0401024,hep-ex-0403011}.  
The measured Gaussian
width of $6.1\pm 1.6({\rm stat.})^{+2.0}_{-1.4}({\rm syst.})\mev$
was above, but rather close to the experimental resolution. 

In order to determine the natural width of the possible state,   
the Breit-Wigner function convoluted with the Gaussian  
was used in the fitting procedure.
If the width of the Gaussian is fixed to the experimental 
resolution, the estimated intrinsic width of the signal was 
$\Gamma=8\pm 4(\mathrm{stat.}) \mev$. The systematical error on this  
width is expected to be smaller than the statistical error, but
due to low statistics, complicated background and the narrowness of the peak
leading to unstable fits, 
full systematical uncertainty was difficult to estimate.

\begin{figure}
\includegraphics[height=6.0cm]{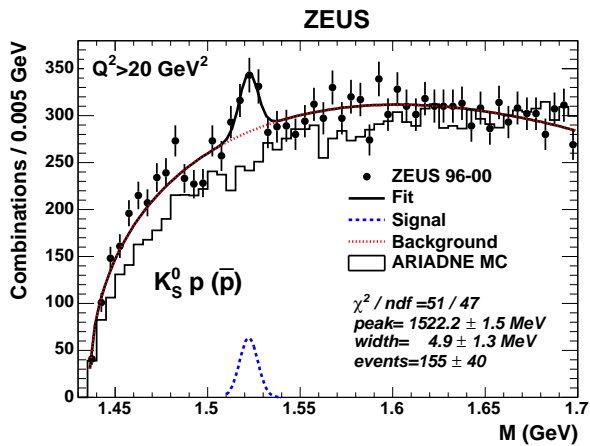}%
\caption{\label{theta1}
Same data as in Fig.~\ref{theta},
but fit is performed using a single Gaussian plus the background.
The $\chi^2/\mathrm{ndf}$ is acceptable since it is dominated
by the region $>1550\mev$. 
For $M<1550\mev$, the $\chi^2/\mathrm{point}$ is $32/22$ for the same fit. 
}
\end{figure}

The number of events ascribed to the signal by this fit was  $221
\pm 48$.  The statistical significance, estimated from the number
of events assigned to the signal by the fit, was  $4.6\sigma$.
Other fit parameters are shown in Table 1. The number of events in
the $\kspb$ channel was $96 \pm 34$. If the signal corresponds to
the pentaquark, this provides the first evidence for its
antiparticle.

%-------------------------------------------------------------------------------
%       An example table
%-------------------------------------------------------------------------------
\begin{table}
\begin{tabular}{|c|c|c|c|c|}
\hline
Fit function & peak & width &  events &  $\chi^2 / \mathrm{ndf}$   \\  \hline \hline
Bkg. & - & - & - &  $69/50$ \\ \hline
Gaussian+Bkg. & $1522.2\pm 1.5$  & $4.9\pm 1.3$  & $155\pm 40$  &  $51/47$ \\ \hline
2 Gaussians+Bkg. & $1521.5\pm 1.5$  & $6.1\pm 1.6$  & $221\pm 48$  &  $35/44$ \\ \hline
\end{tabular}
\caption{\label{ami1} Fit results (in MeV) for $Q^2>20\gev^2$. All
parameters are free in the fitting procedure.}
\end{table}

Under  a conservative assumption, one could assume that 
the enhancement from the left
side of the $1522\mev$ peak has a statistical nature,  or is due to an
instrumental effect which is  not taken into account in the Monte Carlo  
simulation. As seen in Fig.~\ref{theta1}, the single-gaussian fit
plus background is still acceptable according to the $\chi^2/\mathrm{ndf}$
test.  However, in the low-mass region, $M<1550\mev$, this fit had 
a poor quality. 
For the background fit only, the $\chi^2/\mathrm{point}=2.2$,
while the overall $\chi^2/\mathrm{ndf}$, which  is dominated
by the high-mass region,  is reasonable (see Table~1).

It is interesting
that the $\kspb$ data has a peak near $1470\mev$, while the $\ksp$ channel does
not have a clear peak at this mass. 
Generally, one could conclude
that the data may contain a contribution from the unestablished
$\Sigma(1480)$ state. This conclusion is mainly supported by the
comparison of the data with the Monte Carlo  simulation. The parameters and
the significance of this state are difficult to  estimate due to
the steeply falling background.

An ensemble of Monte Carlo experiments was generated  in which the
background shape was parameterised by the threshold function, and the
background rate was simulated as the mean of a Poisson
distribution in each mass bin. The probability of a similar signal
anywhere in the range 1500--1560 MeV arising from statistical fluctuation of
the background was below
$6 \times 10^{-5}$. For a more realistic case, when the background
was parameterised by the threshold function plus the $1465\mev$
Gaussian as the starting distribution, the probability was
found to be about a factor of ten lower.

The mass spectrum was studied over a large $Q^2$ range.
The signal was well seen for $Q^2>60\gev^2$, but, in this case,  
the statistics is
smaller. No significant signal was found at
very low $Q^2$ ($Q^2=1-5\gev^2$) due to a complicated background and
possible acceptance effects. However, the  same signal was found at
$Q^2>1\gev^2$ at low photon-proton centre-of-mass energy, $W<120\gev$.
For this $W$ range, the charged multiplicity 
(and, hence, the combinatorial background) is lower.

In conclusion, the results constitute evidence for the existence a
narrow baryonic resonance decaying to the
$\ksppb$ channel with the mass
near 1522 MeV. This is the first observation of  such  resonance
in high-energy colliding experiments in the phase-space regions
dominated by fragmentation.

The measured mass and the width of the observed state 
are close to those observed in the
$K^+n$ channel \footnote{
In this decay channel, 
the systematical mass-scale uncertainty is usually significant.}, 
and agree well with the theoretical expectations  
\cite{zp:a359:305}. The PDG reports no any $\Sigma$ state in the
invariant-mass region 1480--1560 MeV, and no peak with a similar
mass was observed in $\Lambda\pi$ decays. This favours the
conclusion that the observed peak can be interpreted as a
pentaquark candidate. In this  interpretation, the signal seen in
the $\kspb$ channel corresponds to an
antipentaquark with a quark content of
$\bar{u}\bar{u}\bar{d}\bar{d}s$.

%%%%%%%%%%%%%%%%%%%%%%%%%%%%%%%%%%%%%%%%%%
\section{$\Theta^{++}$ state?}
%%%%%%%%%%%%%%%%%%%%%%%%%%%%%%%%%%%%%%%%%%

\begin{figure}
\includegraphics[height=6.0cm]{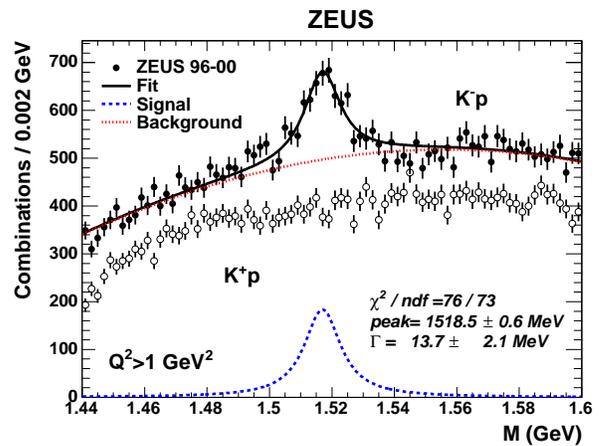}
\caption{\label{theta++}
Invariant-mass spectra for the 
$K^+p$ and $K^-p$ channels (plus charge conjugates) 
for $Q^2 > 1 \gev^2$. }  
\end{figure}

If $\Theta^{+}$ 
is an isotensor state, a
$\Theta^{++}$ signal can  be expected in the $K^+ p$ spectrum
\cite{plb570:185}. The $K^\pm p (K^\pm\bar{p})$ invariant mass
spectra were investigated in  a wide range of minimum $Q^2$
values, identifying proton and charged kaon candidates using
the $dE/dx$ information. 
The proton candidates inside the $dE/dx$ proton band 
were required to have $dE/dx>1.8$ mips,
while the kaon candidates were reconstructed in the kaon band
after the restriction $dE/dx >1.2$ mips. For $Q^2>1\gev^2$, no peak was
observed near 1522 MeV in the $K^+p$ and $K^-\bar{p}$ spectra, 
while a clean signal was seen in
the $K^-p (K^+\bar{p})$ channel at $1518.5\pm 0.6(\mathrm{stat}.)\mev$, corresponding
to the PDG $\Lambda(1520)D_{03}$ state, Fig.~\ref{theta++}.

The numbers of the reconstructed $\Lambda(1520)$ and 
$\bar{\Lambda}(1520)$ candidates agree very well with each other. The same
good agreement was found for $K^+$ and $K^-$, 
as well as for $p$ and $\bar{p}$
candidates. This indicates negligible contributions from the proton
remnant, as well as from possible secondary-scattering events.

\begin{figure}
\includegraphics[height=6.0cm]{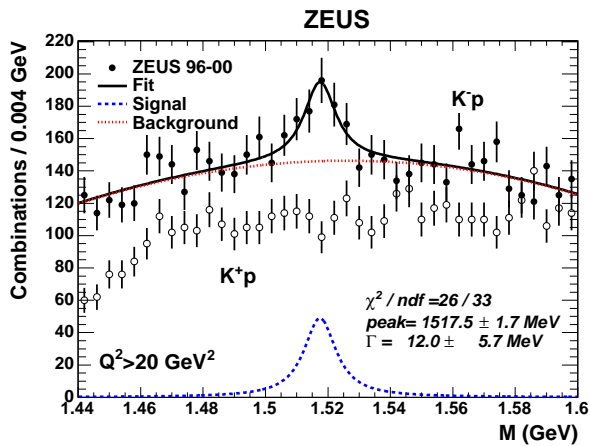}
\caption{\label{theta++1}
Invariant-mass spectra for the $K^+p$ and $K^-p$ channels for $Q^2 >
20 \gev^2$.
}
\end{figure}

The $K^+p$ and $K^-p$ mass spectra were  investigated at
$Q^2>20\gev^2$, i.e. in the kinematic region where the 1522 MeV peak is  
clearly pronounced for the $\ksppb$ channel. Fig.~\ref{theta++1} shows the
mass spectra for $Q^2>20\gev^2$. Again, no sign of a peak was
found in the $K^+p$ decays.

The failure to observe $\Theta^{++}$ indicates that the $\Theta^+$ state
is not isovector or isotensor.

In contrast to the $\ksppb$ state, the detector acceptance for
$\Lambda(1520)$ baryons decreases faster with increase of
$Q^2$, since the momenta of $K^{\pm}$-mesons were restricted by the requirement
$dE/dx>1.2$ mips (i.e. high-momentum mesons were  not used
in the reconstruction), while $\ks$ used 
in the $\ksppb$ reconstruction were not restricted in the total momenta.
Since the detector acceptances  are  significantly
different for the $\ksppb$ and $K^-p$ measurements, direct comparisons of the
production rates of possible pentaquarks and $\Lambda(1520)$ state 
are not possible before taking into account the detector acceptance effects.

% Create the reference section using BibTeX:
\bibliography{penjap}

\end{document}